\documentstyle[preprint,floats,pra,aps,epsfig]{revtex} 
\tightenlines
\begin{document}

\draft

\title{\LARGE \bf Urban tranport phenomena in the street canyon} 

\author{Maciej M. Duras 
\thanks{Electronic address: mduras @ riad.usk.pk.edu.pl}} 
\address{Institute of Physics, Cracow University of Technology, 
ulica Podchor\c{a}\.zych 1, PL-30-084 Cracow, Poland \\
"The Proceedings of the Comphy02 
Computational Physics of Transport and Interface Dynamics conference", 
M. Schreckenberg et al Eds., Springer-Verlag, Berlin, Germany  (2002).} 

\date{AD 2002, 13th March}
  
\maketitle 

\begin{abstract}
A field deterministic model of the vehicular
dynamics in a generic urban street canyon
with two neighboring canyons is considered.
The assumed hydrodynamical model of 
vehicular movement is coupled to the
gasdynamical model of the air and emitted
pollutants. The vehicles are assumed
to move on distinct left lanes and right ones.
At the upstream and downstream ends there are
coordinated traffic lights which introduce
control parameters to the field model.  
The model of optimal control of street canyon dynamics
is based on the two optimal multi-criteria control problems.
The problems consist of minimization of the dimensionless functionals
of the cumulative total travel time,
global emissions of pollutants,
and global concentrations of pollutants,
both in the studied street canyon,
as well as together with the 
two nearest neighbor substitute canyons, respectively.
\end{abstract} 


\section{The field models of vehicles and pollutants.}
\label{sect-field}  
In the present article we consider two coupled deterministic
field model of the urban street canyon.
The vehicular fields are one-dimensional in spatial
variable $x$ and they depend on time variable $t$.
The considered fields are the vehicular number density $k_{l, vt}^{s}(x, t)$,
vehicular velocity ${\bf{w}}_{l, vt}^{s}(x, t)$,
emissivity of exhaust gases $e_{l, ct, vt}^{s}(x, t)$,
and vehicular heat emissivity $\sigma_{l, vt}^{s}(x, t)$,
where $vt$ is emission type's number,
$ct$ is number of the constituent of emitted exhaust gases.
It is assumed that there are $n_{1}=n_{L}$ left lanes 
$s=1$ and $l$ is the left lane's number, 
$l=1, ..., n_{L},$
whereas 
for $n_{2}=n_{R}$ right lanes $s=2$ and $l$ is the right lane's number,
$l=1, ..., n_{R},$ 
$vt=1, ..., VT$,
$ct=1, ..., CT$.
$VT$ is total number of types of vehicular emissions,
and $CT$ is total number of emitted exhaust gases.
The abovementioned model consists of the equations 
of balances of vehicular numbers \cite{Michalopoulos 1984}
and vehicular equations of state 
(Greenshields' equilibrium speed-density u-k model
\cite{Greenshields 1934})
as well as exhaust gas emissivities and heat emissivities.
The vehicular flow in the canyon is multilane bidirectional one-level
rectilinear, and it is considered
with two systems of coordinated signalized junctions
at the upstream and downstream ends of lanes
\cite{Duras 1999 engtrans numer},\cite{Duras 1998 thesis}, 
\cite{Duras 1997 PJES}, \cite{Duras 1999 PJES}.
The considered vehicles belong to distinct vehicular classes:
passenger cars, and trucks.
Emissions from the vehicles are based on technical measurements
of the following types of pollutants:
carbon monoxide CO, hydrocarbons HC, nitrogen oxides ${\rm NO_{x}}$.
The field model of pollutants is gasdynamical.
The gases are Newtonian viscous perfect and
noninteracting.
The considered fields are three-dimensional in spatial
variables $(x, y, z)$ and one-dimensional in temporal
variable $t$ and they are mixture's density $\rho(x, y, z, t)$,
velocity ${\bf{v}}(x, y, z, t)$, 
temperature $T(x, y, z, t)$, pressure $p(x, y, z, t)$, 
as well as constituents'
mass concentrations $c_{i}(x, y, z, t)$,
and partial pressures $p_{i}(x, y, z, t),i=1, ..., N$,
where $N$ is total number of mixture's elements.
($N=N_{E}-1 +N_{A}$). 
The first $N_{E}-1=3$ gases are the exhaust gases emitted by  
vehicle engines during combustion 
${\rm CO, CH, NO_{x}}$. 
The remaining $N_{A}=9$ gases are the constituents of air: 
${\rm O_{2}, N_{2}, Ar, CO_{2}, Ne, He, Kr, Xe, H_{2}}$. 
The field equations are the balances of
mixture's mass, linear momentum, energy,
and equation of state (Clapeyron's law) as well as
the balances of masses of constituents and
constituents' equations of states (Dalton's law).
The studied sources are mixture mass source $S(x, y, z, t)$,
mixture energy source $\sigma(x, y, z, t)$,
and constituents' mass sources $S^{E}_{i}(x, y, z, t)$
(exhaust gases and consumed oxygen).
The boundary-initial problems are of mixed types
(Dirichlet's and von Neumann's types).
The equations of dynamics are solved by the finite difference
scheme.
The two separate multi-criteria optimization problems are posed
by defining the dimensionless functionals of 
cumulative total travel time, global emissions of pollutants, 
and global concentrations of pollutants,
either in the studied street canyon
or in the canyon and its two nearest neighbor substitute canyons.
The vector of control is a five-tuple
composed of two cycle times, two green times,
and one offset time between the traffic lights.
The optimal control problems consists of
minimization of the two functionals
over the admissible control domain manifold.

\section{The governing equations.}
\label{sect-dynamics} 
Under the abovementioned specifications, 
the set of governing equations is formulated as follows
\cite{Eringen 1990}, \cite{Landau 1986}:
\begin{eqnarray}
& & \frac{\partial k_{l, vt}^{s}}{\partial t} 
+ {\rm div}(k_{l, vt}^{s}{\bf{w}}_{l, vt}^{s})=0.  
\label{vehicle-continuity-eq-a} \\
& & {\bf{w}}_{l, vt}^{s}(x, t)=
(
w_{l, vt, f}^{s} \cdot
(1-\frac{k_{l, vt}^{s}(x, t)}{k_{l, vt, {\rm jam}}^{s}}), 0, 0).
\label{Greenshields-eq-a} \\
& & \frac{\partial \rho}{\partial t}
+ {\rm div}(\rho {\bf{v}})=S.
\label{continuity-eq} \\
& & \rho (\frac{\partial {\bf{v}}}{\partial t}
+ ({\bf{v}} \circ \nabla) {\bf{v}}) + S{\bf{v}}=
- \nabla p + \eta \Delta {\bf{v}} 
+ (\xi + \frac{\eta}{3}) \nabla ({\rm div} {\bf{v}}) + {\bf{F}}.
\label{Navier-Stokes-eq} \\
& & \rho (\frac{\partial \epsilon}{\partial t} + {\bf{v}} \circ \nabla \epsilon)=
-(-\frac{1}{2}{\bf{v}}^{2} + \epsilon) S
+ {\bf T}:\nabla {\bf{v}}+{\rm div}(-{\bf{q}})+\sigma.
\label{energy-eq} \\
& & \frac{p}{\rho}=\frac{R}{m_{{\rm air}}} \cdot T.
\label{state-eq} \\
& & \rho (\frac{\partial c_{i}}{\partial t} + {\bf{v}} \circ \nabla c_{i})=
S^{E}_{i} - c_{i} S + 
\label{diffusion-eq-a} \\
& & + \sum_{m=1}^{N-1}\{ (D_{im}-D_{iN})
\cdot {\rm div}[\rho \nabla (c_{m}+\frac{k_{T, m}}{T} \nabla T)] \}. 
\nonumber \\
& & p_{i}=c_{i} \cdot \frac{m_{{\rm air}}}{m_{i}} \cdot p,
\label{constituent-state-eq}
\end{eqnarray}
where 
$w_{l, vt, f}^{s}$ is maximum free flow speed, 
$k_{l, vt, {\rm jam}}^{s}$ is jam vehicular density, 
$\eta$ is the first viscosity coefficient for the air, 
$\xi$ is the second viscosity coefficient, 
${\bf{F}}=\rho {\bf{g}}$ is the gravitational body force density,
${\bf{g}}$ is the gravitational acceleration,
$D_{im}=D_{mi}$ is the mutual diffusivity coefficient 
from the $i$-th constituent to $m$-th one, and $D_{ii}$  is the 
autodiffusivity coefficient of the $i$-th constituent,
and $k_{T, m}$ is the thermodiffusion ratio of the $m$-th constituent,
$\epsilon$ is the mass density of intrinsic (internal) energy 
of the air mixture, 
${\bf T}$ is the stress tensor, 
symbol $:$ denotes the contraction operation,
$\bf{q}$ is the vector of flux of heat.
We assume that \cite{Duras 1998 thesis}:
\begin{eqnarray} 
& & \epsilon=\sum_{i=1}^{N}\epsilon_{i},
\label{intrinsic-energy-def} \\
& & \epsilon_{i}=
\frac{1}{m_{i}} 
\{ 
c_{i} k_{B} T 
\exp(-\frac{m_{i}|{\bf{g}}|z}{k_{B}T}) \cdot
[
(-\frac{z}{c}) \cdot (1-\exp(-\frac{m_{i}|{\bf{g}}|c}{k_{B}T}))
-\exp(-\frac{m_{i}|{\bf{g}}|c}{k_{B}T})
]
\} +
\label{ith-intrinsic-energy-def} 
\nonumber \\
& & + \tilde{\mu}_{i}c_{i}, 
\nonumber \\
& &
\tilde{\mu}_{i}=\frac{\mu_{i}}{m_{i}},
\label{chemical-potential-tilde} \\
& & 
\mu_{i}= 
k_{B} T \cdot
\{
\ln
[
(c_{i} p) (k_{B} T)^{-\frac{c_{p, i}}{k_{B}}}
(\frac{m_{{\rm air}}}{m_{i}}) (\frac{2\pi h^{2}}{m_{i}})^{\frac{3}{2}}
]
\}
+m_{i} |{\bf{g}}| z
,
\label{chemical-potential-def} \\
& & {\rm T}_{mk}=-p\delta_{mk}
+
\label{stress-tensor-def} \\
& & + \eta \cdot 
[
(
\frac{\partial v_{m}}{\partial x_{k}}
+
\frac{\partial v_{k}}{\partial x_{m}}
-
\frac{2}{3} \delta_{mk} {\rm div}({\bf{v}})
)
\frac{\partial v_{k}}{\partial x_{m}}
]
+
\xi \cdot
[
(
\delta_{mk} {\rm div}({\bf{v}})
)^{2}
],
m, k=1, ...,3,
\nonumber \\
& & {\bf T}:\nabla {\bf{v}}=
\sum_{m=1}^{3} \sum_{k=1}^{3} 
{\rm T}_{mk} \frac{\partial v_{m}}{\partial x_{k}}, 
\label{stress-tensor-velocity-contraction} \\
& & {\bf{q}}=
\sum_{i=1}^{N}
\{
[
(\frac{\beta_{i}T}{\alpha_{ii}}
+ \tilde{\mu}_{i}) {\bf{j}}_{i}
]
+
[
(-\kappa) \nabla T
]
\}
,
\label{vector-flux-heat-def} \\
& & {\bf{j}}_{i}=
-\rho D_{ii} (c_{i}+\frac{k_{T, i}}{T} \nabla T)
,
\label{vector-flux-mass-def} \\
& & \alpha_{ii}=
\frac{
[\rho D_{ii}]
}
{ 
[(\frac{\partial \tilde{\mu}_{i}}{\partial c_{i}})
_{(c_{n})_{n=1, ..., N, i \neq n}, T, p}]
},
\label{alpha-def} \\
& & \beta_{i}=
[\rho D_{ii}] \cdot
\{
\frac{k_{T, i}}{T}
-
\frac{
[(\frac{\partial \tilde{\mu}_{i}}{\partial T})
_{(c_{n})_{n=1, ..., N}, p}]
}
{
[(\frac{\partial \tilde{\mu}_{i}}{\partial c_{i}})
_{(c_{n})_{n=1, ..., N, i \neq n}, T, p}]
}
\}
,
\label{beta-def}
\end{eqnarray}
where 
$\epsilon_{i}$ is the mass density of intrinsic (internal) energy 
of the $i$th constituent of the air mixture,
$m_{i}$ the molecular mass of the $i$th constituent,
$k_{B}$ 
is Boltzmann's constant,
$\mu_{i}$ is the complete partial chemical potential of the $i$th constituent 
of the air mixture (it is complete since it is composed of
chemical potential without external force 
field and of external potential),
$m_{{\rm air}}=28.966$ [u] is the molecular mass of air,
$\delta_{mk}$ is Kronecker's delta,
$c_{p, i}$ is the specific heat at constant pressure of the $i$th 
constituent of air mixture,
$h$ is Planck's constant,
${\bf{j}}_{i}$ is the vector of flux of mass of the $i$th constituent
of the air mixture,
and $\kappa$ is the coefficient of thermal conductivity of air.
These magnitudes were derived from
Grand Canonical ensemble with external gravitational Newtonian field.

\section{\bf Optimal control problems.}
\label{sect-optimization}  
Let us consider the measures 
of the total travel time (TTT): $J_{{\rm TTT}}$ 
\cite{Michalopoulos 1984},
emissions (E): $J_{{\rm E}}$ , and concentrations (C): $J_{{\rm C}}$ 
of exhaust gases in the street canyon
as well as in the canyon and two its nearest neighbor substitute canyons,
$J_{{\rm TTT, ext}}$ $J_{{\rm E, ext}}$ $J_{{\rm C, ext}}$, respectively. 
Therefore the appropriate optimization problems 
may be formulated as follows \cite{Duras 1998 thesis}.
The measures depend on traffic lights at the
upstream and downstream ends of the lanes.
The vector of boundary control ${\bf u}$ reads:
\begin{equation}
{\bf {u}}=(g_{1}, C_{1}, g_{2}, C_{2}, F) \in U^{{\rm adm}},
\label{control-5-tuple-eq}
\end{equation}
$g_{m}$ are green times, 
$C_{m}$ are cycle times,
$F$ is offset time, 
and $U^{{\rm adm}}$ is a set of admissible control variables. 
The TTT, E, C functionals are given by:
\begin{equation}
J_{{\rm TTT}}({\bf{u}}) 
=\sum_{s=1}^{2} \sum_{l=1}^{n_{s}} \sum_{vt=1}^{VT}
\int_{0}^{a} \int_{0}^{T_{S}}
k_{l, vt}^{s}(x, t) dx \, dt .
\label{functional-TTT-def} 
\end{equation}
\begin{equation}
J_{{\rm E}}({\bf{u}})
=\sum_{s=1}^{2} \sum_{l=1}^{n_{s}} \sum_{ct=1}^{CT} \sum_{vt=1}^{VT}
\int_{0}^{a} \int_{0}^{T_{S}}
e_{l, ct, vt}^{s}(x, t) dx \, dt .
\label{functional-E-def} 
\end{equation}
\begin{equation}
J_{{\rm C}}({\bf{u}})=
\rho_{{\rm STP}} \cdot 
\sum_{i=1}^{N_{E}-1} 
\int_{0}^{a} \int_{0}^{b} \int_{0}^{c} \int_{0}^{T_{S}}
c_{i}(x, y, z, t) dx \, dy \, dz \, dt.
\label{functional-C-def}
\end{equation}
\begin{eqnarray}
& & J_{{\rm TTT, ext}}({\bf{u}})=
J_{{\rm TTT}}({\bf{u}}) + 
\label{functional-TTT-ext-def} \\
& & +
a \cdot \sum_{s=1}^{2}
\sum_{l=1}^{n_{s}} \sum_{vt=1}^{VT} k_{l, vt, {\rm jam}}^{s}
\cdot (C_{s}-g_{s}).
\nonumber
\end{eqnarray}
\begin{eqnarray}
& & J_{{\rm E, ext}}({\bf{u}})=
J_{{\rm E}}({\bf{u}}) +
\label{functional-E-ext-def} \\
& & +
a \cdot \sum_{s=1}^{2}
\sum_{l=1}^{n_{s}} \sum_{ct=1}^{CT} \sum_{vt=1}^{VT} 
e_{l, ct, vt, {\rm jam}}^{s}
\cdot (C_{s}-g_{s}).
\nonumber 
\end{eqnarray}
\begin{eqnarray}
& & J_{{\rm C, ext}}({\bf{u}})=
J_{{\rm C}}({\bf{u}}) +
\label{functional-C-ext-def} \\
& & +
\rho_{{\rm STP}} \cdot a \cdot b \cdot c \cdot 
\sum_{i=1}^{N_{E}-1} c_{i, {\rm STP}} \cdot 
\sum_{s=1}^{2} (C_{s}-g_{s}).
\nonumber
\end{eqnarray}
The integrands $k_{l, vt}^{s}, 
e_{l, ct, vt}^{s}, c_{i}$ 
in the functionals depend on the control vector ${\bf u}$ 
through the boundary conditions,
through the equations of dynamics,
as well as, through the sources.
The value of the vector of control $u$ directly
affects the boundary conditions
for vehicular densities, velocities, and emissivities.
It also affects the sources.
Next, it propagates to the equations of dynamics
and then it influences the values of functionals.
$\rho_{{\rm STP}}$ is the density of air 
at standard temperature and pressure STP,
$c_{i, {\rm STP}}$ is concentration of the $i$th
constituent of air at standard temperature and pressure,
$a, b, c$, are dimensions of the street canyon,
$T_{S}$ is time of simulation.

We define two additional functionals.
The first one is the composite total travel time,
emissions and concentrations of the pollutants in the single
canyon, whereas the second one in the canyon and
two its neighbors:
\begin{eqnarray}
& & H({\bf{a}}, {\bf{u}}) 
= \alpha_{{\rm TTT}} F_{{\rm TTT}}({\bf{u}}) 
+ \alpha_{{\rm E}} F_{{\rm E}}({\bf{u}}) 
+ \alpha_{{\rm C}} F_{{\rm C}}({\bf{u}}).
\label{functional-H-def} \\
& & H_{{\rm ext}}({\bf a}, {\bf{u}})=
H({\bf a},{\bf{u}}) + 
\label{functional-H-ext-def} \\
& & + \alpha_{{\rm TTT, ext}} J_{{\rm TTT, ext}}({\bf{u}}) 
+ \alpha_{{\rm E, ext}} J_{{\rm E, ext}}({\bf{u}}) 
+ \alpha_{{\rm C, ext}} J_{{\rm C, ext}}({\bf{u}}).
\nonumber
\end{eqnarray}
The scaling parameters 
$\alpha_{{\rm TTT}},\alpha_{{\rm E}},\alpha_{{\rm C}},
\alpha_{{\rm TTT, ext}},\alpha_{{\rm E, ext}},\alpha_{{\rm C, ext}}$
make the two functionals dimensionless, and 
\begin{eqnarray}
& & {\bf a}=(\alpha_{{\rm TTT}},\alpha_{{\rm E}},\alpha_{{\rm C}}) 
\in A^{{\rm adm}},
\label{a-def} \\
& &{\bf a}_{{\rm ext}}=(\alpha_{{\rm TTT}},\alpha_{{\rm E}},\alpha_{{\rm C}},
\alpha_{{\rm TTT, ext}},\alpha_{{\rm E, ext}},\alpha_{{\rm C, ext}})
\in A_{{\rm ext}}^{{\rm adm}},
\label{a-ext-def}
\end{eqnarray} 
where $A^{{\rm adm}}$, $A_{{\rm ext}}^{{\rm adm}}$ 
are the sets of admissible scaling parameters.
We formulate two separate multi-criteria optimization
problems consisting in minimization
of the functionals $H, H_{{\rm ext}}$
with respect to control vector ${\bf u}$ over admissible domain
and with respect to scaling parameters, while the equations
of dynamics are fulfilled: 
\begin{equation}
H^{\star}=H({\bf{a}}^{\star}, {\bf{u}}^{\star})=
\min \{ {\bf{a}}\in A^{{\rm adm}}, {\bf{u}}\in U^{{\rm adm}}: 
H({\bf{a}}, {\bf{u}}) \}.
\label{optimum-H-def}
\end{equation}   

\begin{equation}
H_{{\rm ext}}^{\star}=
H_{{\rm ext}}({\bf{a}}_{{\rm ext}}^{\star}, {\bf{u}}_{{\rm ext}}^{\star})=
\min \{ {\bf{a}}_{{\rm ext}} \in A_{{\rm ext}}^{{\rm adm}}, 
{\bf{u}} \in U^{{\rm adm}}: 
H_{{\rm ext}}({\bf{a}}, {\bf{u}}) \},
\label{optimum-H-ext-def}
\end{equation}   
where $H^{\star}, H_{{\rm ext}}^{\star}$,
are the minimal values of the two functionals,
$({\bf{a}}^{\star}, {\bf{u}}^{\star})$, 
and $({\bf{a}}_{{\rm ext}}^{\star}, {\bf{u}}_{{\rm ext}}^{\star})$,
are scaling and control vectors
at which the functionals reach the minima, respectively.

\section{Acknowledgements.}

The author would like to gratefully thank
the organizers of the 
Comphy02 Computational Physics of Transport and Interface Dynamics
conference:
Prof. Dr. Heike Emmerich, Britta Nestler, and 
Michael Schreckenberg,
held in the Max-Planck-Institut f\"ur Physik
komplexer Systeme, Dresden, Germany, 
for the invitation and creation of scientific atmosphere.
The present work was partly covered by the 
Polish State Commitee for Sientific Research 
KBN grant number F-1/63/BW/02.

\section{Conclusions.}
\label{sect-conclusions}

The proecological urban traffic control idea 
and the model of the street canyon have been developed. 
It is assumed that the proposed model
represents the main features of complex air pollution phenomena.

\end{document}